\documentclass{article}

\usepackage{url}
\usepackage{listings}
\usepackage{moreverb}
\usepackage{a4wide}

\begin{document}

\title{Determine the User Country of a Tweet}
\author{Han van der Veen, Djoerd Hiemstra, Tijs van den Broek,\\
Michel Ehrenhard, and Ariana Need\\ \vspace{0.1cm}
University of Twente}

\date{}
\maketitle

\begin{abstract}
In the widely used message platform Twitter, about 2\% of the tweets contains the geographical location through exact GPS coordinates (latitude and longitude). Knowing the location of a tweet is useful for many data analytics questions. This research is looking at the determination of a location for tweets that do not contain GPS coordinates. An accuracy of 82\% was achieved using a Naive Bayes model trained on features such as the users' timezone, the user's language, and the parsed user location. The classifier performs well on active Twitter countries such as the Netherlands and United Kingdom. An analysis of errors made by the classifier shows that mistakes were made due to limited information and shared properties between countries such as shared timezone. A feature analysis was performed in order to see the effect of different features. The features \emph{timezone} and \emph{parsed user location} were the most informative features.
\end{abstract}

\section{Introduction}
Twitter is a micro-blogging network that connects 284 million active users and produces approximately 500 million tweets everyday~\cite{twitter2014about}. On this platform users are able to send messages in 140 character tweets to their followers worldwide. The country from which a tweet was end is not directly available in the tweet. However, this information can be deduced by combining all available data.

This research is conducted by making use of the Twitter datagrant\cite{twitter2014datagrant}. University of Twente has received access to all tweets related to the topic of \emph{The Diffusion And Effectiveness of Cancer Early Detection Campaigns on Twitter}. Our team will study \emph{cancer awareness campaigns} by examining the effects of tweets about specific campaigns, such as Movember. The aim of the Movember campaign is to raise awareness for men's health and raise funds for cancer research. Our project aims to compare Movember national campaigns by analyzing Twitter messages. As an important step towards cross-country analysis. This paper will investigate how to identify the country of a tweet.

The \emph{country of a tweet} is the country where the user of the tweet currently lives. The country from which the tweet itself was send can be different. For example, a user can tweet while on holiday in Paris, but live in New York. We will examine the origin of a tweet by determining the country of the user. To do so we use (meta)information of a tweet itself, for example the language of a tweet. Additionally, we use Geocoding services to find a location with just a query string. Geocoding can be difficult due to ambiguous location names such as \emph{Nieuw Amsterdam} in the Netherlands or \emph{New Amsterdam} in the United States. This research combines an existing Geocoding service with a machine learning model to improve the accuracy. We will evaluate our method against the GPS coordinates in a Tweet. For this problem standard classification models can be applied, such as Na\"ive bayes or Support Vector Machines.

This research is aimed at scientists who want to determine the country of a tweet. However, it can also be used more generally to classify tweets into certain classes. The classes can be, for instance, the language of the tweet or gender of a tweet user. Another application might be a method for filtering a set of tweets into a different classes.

A location of a tweet can be provided by adding coordinates or location to a Tweet. A tweet location is expressed in a latitude and a longitude. With these coordinates a country can easily be extracted by using a Geocoding service, such as Google Maps Geocoding API\footnote{https://developers.google.com/maps/documentation/geocoding/}. The problem is that about 2\% (see section~\ref{sec:results}) of the tweets has enabled this feature. To determine the country for all tweets we have to establish a method to use other available (meta-)information to identify the country of a tweet. 

Hence, \textbf{our research investigates which Tweet features are required to identify the country of a single tweet}.

\section{Related work}
In Twitter some terms are used that are used a lot. For example, a hashtag is a keyword that starts with a \# and identifies a topic the tweet is related to. A tweet can also contain a link or a location. For example, a tweet posted from a basketball match. A retweet is a copy of a tweet that is posted to your own wall. A retweet can be used to share messages. 

Twitter and other social networks consists of several building blocks~\cite{kietzmann2011social}. These blocks are divided into 7 groups: Presence, Sharing, Relationships, Conversations, Reputation, Groups and most importantly Identity. In social networks all these blocks are available. In Twitter the blocks Sharing, Conversations and Reputation are most important. 

Every tweet has some impact for the user. Kwak et al.~\cite{kwak2010twitter} have measured the following characteristics of Twitter. There are few users which have a lot of followers and on the contrary there are many users which have only a few followers. When a users has more than 100 followers, then they have more than 1000 tweets. However, when a user reaches more than 10000 followers the amount of tweets are not evenly distributed anymore. Twitter users will mostly read tweets of users that have the same timezone, same country or city. There are few people who are really influencing the twitter world. Some users have a reach of more than 10000 people. But even small users can reach up to 1000 users using retweets.

The Tweets are freely available to any developer~\cite{twitter2014developertweets}. In each tweet the information about the user is included and all the information of the message itself, such as create\_date, user, list of hashtags, language, geolocation, some statistics regarding the tweet is favorited and more. All this information is available through the Twitter API. We can search the Twitter API for a certain hashtag, for example \#movember.

\subsection{Determining a location using Twitter data}
There are several approaches to determine the location of a tweet user. One can look at the tweets text, examine meta-information or look at the relationships of users.

Chengh et al.~\cite{cheng2010you} used tweet text to determine the location. However, tweet texts have a lot of noise and people may travel and tweet about different locations. Changt et al. are only using tweet within the United States. Their classifier is using the function \emph{ErrorDistance} for estimating the distance between the real city and the determined location from the words. To do so, they are using TF/IDF indexing and a smoothing function to optimize the result. Which results in an average error of about 500 miles. Hecht et al.~\cite{hecht2011tweets} has the same approach. They are using a Bayes classifier for words, that is optimized using the extra condition that require that a word is at least used once by a certain amount of users in that class. When this is not the case the probability of a word is zero, in all other cases the Bayes function is used. Their accuracy is around 74\% using a uniform dataset.

Priedhorsky et al.~\cite{priedhorsky2014inferring} used tweet information as their data and with a Gaussian mixture model they classify not only the country but the city of a tweet as well. By minimizing the error of the Gaussian Mixture model they have build their classifier. They have build n-grams of the tweet text and using the n-grams as features. They show that adding time zone will improve the accuracy rate with 3\% and adding user location only 1.3\%. Tweet text will improve the accuracy with 7\%. 

Zhang et al.~\cite{zhang2014geocoding} propose a method that also uses meta-information in their classifier. Their main focus is to be able to extract the proper location from ambiguous locations. A list of toponyms are used to annotate the tweets and used to parse the tweets. When geoparsing a tweet, a list of candidates is ranked with a Support Vector Machine. Also features such as timezone and user location are added to the classifier in order to improve the accuracy. Using a list of geo location names, such as cities and countries, and meta-information, their classifier reached an accuracy of 84\%.

Another method of detecting locations in text is using entity recognition software. Gelernter et al.~\cite{gelernter2011geo} are using entity recognition software. The reached accuracy was low, mainly due to spelling mistakes and abbreviations of the same entity, which were not known. Manual annotators had an accuracy of 72\%. However, their original method was still useful for mapping tweets on a a map.

Mahmud et al.~\cite{mahmud2014home} determined the home location by looking at all the available information of a user. They combined tweet behavior and tweet meta-information. They are also looking at foursquare data which can be included in a tweet. Foursquare links can be used to extract the location of a tweet. Three features were used to classify tweets. Firstly, a content based feature. Secondly, a hashtag based feature and lastly, extraction of place names feature. A Bayes classifier combines this features into a result. Resulting in a accuracy of 73\%.

Determining a location using followers of users is studied by Davis et al.~\cite{davis2011inferring}. Looking at the tweets of followers of following which have geo location available they are determined the location of that user. When a user is active and has a lot of active friends the accuracy was about 90\%. However, that was only the case for about 40\% of the users in their test set. Chandra et al.~\cite{chandra2011estimating} used user conversations, instead of the connections. When in a conversation geo information is available, a conversation will be linked to that location. This method however resulted in an accuracy of 50\% with an average error distance of 300 miles.

Literature shows that (meta)information will improve most classifiers. Also tweet text improves the accuracy. 

\section{Research Method}
In order to establish the accuracy for this research we need to state how we evaluate. Furthermore, we need to describe precisely which model is used and how the evaluation is applied.

\subsection{Evaluation}
To calculate the accuracy of our classifier, we need to have an annotated set of tweets. We did not do this by hand, but we automatically annotated the tweets and add it into a certain class under the following assumption. 

\textbf{Definition 1.}~
A country of a tweet user can be determined with the geolocation of a tweet.

We have two datasets. Both datasets are extracted from a set that was made available to the datagrant team by Twitter. The dataset consist of about a year of tweets with the word \emph{cancer}\footnote{Also cancer in different languages was searched for} in it. All these tweets were put into classes corresponding to their geolocation coordinates. For instance, a tweet posted in New York will be classified in the class \emph{United States}. We argue that this assumption is acceptable, because most of tweets with geolocation have the geo functionality enabled for the whole account. Furthermore, people are likely to stay most of their time in the same country. However, with enough tweets that will possibly not affect the results, because there are enough tweets to compensate the errors. 

\begin{description}
\item[dataset \#1] 39293 tweets containing \emph{movember}. These tweets were extracted out of 1.5 million tweets about movember. All these tweets have a geolocation.
\item[dataset \#2] 4477 tweets. Randomly extracted from the datagrant cancer data set. All these tweets have a geolocation.
\end{description}

To calculate the accuracy on a data set, 10-fold-cross-validation is applied to both datasets. Accordingly, the evaluation runs 10 times with each time a different 10\% sample of the data as training data and evaluated these with the other 90\%. This will produce an averaged accuracy on the whole dataset. The accuracy of the evaluation will be the average of all iterations. The accuracy is calculated by the amount of tweets where the geolocation country (class) is the same as determined by the Bayes classifier ($T_{same}$). This amount is divided by the total amount of tweets in the dataset ($T_n$). 

\begin{equation}
Accuracy = \frac{T_{same}}{T_n}
\end{equation}

\subsection{Model}
Tweets are just texts with meta-information attached to them, so it might be useful to examine classic classification models. Manning et al.~\cite{manning2008introduction} discussed several classification models. The most used models are Na\"ive Bayes, Support Vector Machines, Decision Trees and Neural Networks. Each model has their advantages and disadvantages. This paper will use the Na\"ive Bayes model. Na\"ive Bayes is great for noisy data and performs well when there are a few classes used. Also combining features into a Bayes model is easy and consequently, each feature can improve the accuracy. 

Our Bayes model is derived from the Bayes model described in Manning et al.~\cite{manning2008introduction}. Instead of counting words we are counting features of a tweet. The tweet fields we will use as features for the Bayes model are: \emph{timezone, user location, tweet language, utc\_offset and geoparsed user location}. When a field is empty we ignore it, because it holds no indication for any country. The \emph{utc\_offset} holds the timezone offset to Greenwich Mean Time in seconds. For example, Netherlands has +1 hour, thus fields \emph{utc\_offset} will be 3600. The field \emph{timezone} contains a place in the current timezone, for example \emph{Amsterdam} or \emph{Paris}. The \emph{tweet language} is determined by twitter itself, probably also with a machine learned classifier, such as Bayes classifier. This field contains a language code using BCP47\footnote{http://tools.ietf.org/html/bcp47} notation. The field \emph{user language} chosen by the user itself and corresponds to the language the user wants their user interface.


For training Bayes, we use the following geo information fields: \emph{geo, coordinates\footnote{Coordinates has the format described in \url{http://geojson.org/geojson-spec.html}} and place}. Geo and coordinates contains the same information, but the field place contains an exact link to a place like Eiffel Tower. The geo information is converted to a country with the HERE geocoding service\footnote{Nokia HERE maps \url{https://developer.here.com/}}, using the coordinates available in the geo information. The features are trained with these countries, under the assumption of definition~1. As an example we will use the tweet in Figure~\ref{fig:tweetjson}. The class of the tweet is Netherlands using geo information. In the class Netherlands we will add 1 to value \emph{Amsterdam} in feature \emph{timezone}. Add 1 to value \emph{Awesome Enschede} in feature \emph{user location} and so on for all features. After counting, the Bayes classifier will learn the probability of each occurrence in a class. For example, Amsterdam will probably have a high count in the class Netherlands. The probability is calculated by dividing the Amsterdam count by the total count in the class Netherlands. 

\begin{figure}[htb]
\centering
\begin{verbatimtab}[4]
{
	created_at: "Wed Jul 24 14:45:20 +0000 2013",
	text: "OH: "Misschien kunnen we iets gaan doen
	met Movember? Wanneer is dat ook alweer?"",
	user: {
		name: "Han van der Veen",
		screen_name: "Haneev",
		location: "Awesome Enschede",
		created_at: "Tue Jun 10 10:46:43 2008",
		utc_offset: 3600,
		time_zone: "Amsterdam",
		geo_enabled: true,
		statuses_count: 20182,
		lang: "nl"
	},
	geo: ...
	coordinates: {
		type: "Point",
		coordinates: [
			4.48431747,
			52.1674388
		]
	},
	place: {
		id: "99ad54d1cccb950b",
		place_type: "city",
		name: "Leiden",
		full_name: "Leiden",
		country_code: "NL",
		country: "Nederland",
		bounding_box: ...
	},   
	lang: "nl"
}
\end{verbatimtab}
\caption{Incomplete example of a Tweet in JSON}
\label{fig:tweetjson}
\end{figure}

The \emph{geoparsed user location} is determined by putting the content of the user location field into a geocoding service, such as Google maps. The returned country is added as a feature into the model. For example, when a user has ''Amazing New York'' as user location, the service will produce \emph{United States}. 

Because we are using Na\"ive Bayes, we assume that all features are independent. The features \emph{timezone} and \emph{utc\_offset}, and \emph{user location} and \emph{geoparsed user location}, are obviously not independent, because \emph{timezone} and \emph{utc\_offset} are different notations for the same timezone. Also \emph{user location} and \emph{geoparsed user location} are related, because the \emph{user location} is only converted into another format. In section~\ref{sec:results} each feature is independently tested to see the impact of violating this assumption.

\section{Results}
\label{sec:results}
We have used the following fields of a tweet: \emph{timezone, location, tweet language and geoparsed user location}. Each of these features was counted and used for classifying the tweets. However, timezone and location were not always available. According to Hecht et al.~\cite{hecht2011tweets} 18\% of the tweets user location field is empty. In 16\% the user typed non-geographical information, such as \emph{everywhere}. In about 66\% of the users \emph{user location} does contain useful information. In 25\% of the tweets the timezone is empty. When a timezone is not empty means that the user has declared that he or she lives in that timezone.  

In our test set there are 128 different \emph{timezones} names identified. There were 13327 different locations and 28 different languages. 

The \emph{timezone} and \emph{tweet language} have data that is limited by Twitter. However, for the location feature the user can type anything in their profile. For example, \emph{on the moon} or \emph{everywhere} are common locations. Fortunately, a lot of people do fill in something useful. Such as, \emph{Amsterdam, NL} or \emph{Enschede}. Because we are using a Bayes classifier, the location \emph{everywhere} can also indicates that the user lives in a English country. Consequently, in the Netherlands we would use the dutch word \emph{overal} which indicate that we are living in the Netherlands.

We have applied 10-fold-cross-matching on the dataset and that resulted in an accuracy of about 82.05\% using the best feature combination for dataset \#1. In table~\ref{tab:features} the features are evaluated separately. Dataset \#1 was 10 times larger than dataset \#2. We found some indication that more training data produces higher accuracy. This is not the case for feature \emph{user language}, for all other features more data results in higher accuracy. 

\begin{table}[t]
\centering
\small
\begin{tabular}[c]{|c|c|c|c|c|c|c|c|}
\hline
~\textbf{Loc}~ & \textbf{$\!$Timezone$\!$} & \textbf{$\!$Language$\!$} & \textbf{$\!$Geoparsed$\!$} & \textbf{UTC} & \textbf{$\!\!$User language}$\!\!$ & \textbf{\#1} & \textbf{\#2} \\
\hline x & & & & & & ~56.18\%~ & 44.21\% \\
\hline & x & & & & & 65.47\% & 62.84\% \\
\hline & & x & & & & 44.02\% & 56.43\% \\
\hline & & & x & & & 73.49\% & 65.77\% \\
\hline & & & & x & & 58.49\% & 57.12\% \\
\hline & & & & & x & 43.92\% & 57.04\% \\
\hline x & & & x & & & 75.14\% & 65.88\% \\
\hline x & x & & & & & 73.08\% & 63.92\% \\
\hline x & x & x & & & & 71.60\% & 67.34\% \\
\hline & x & & x & & & 80.96\% & 73.01\% \\
\hline & & x & x & & & 71.80\% & 68.47\% \\
\hline x & x & x & x & & & 80.84\% & 74.40\% \\
\hline x & x & & x & & & \textbf{82.05}\% & 73.21\% \\
\hline x & x & x & x & x & x & 77.98\% & \textbf{74.69}\% \\
\hline
\end{tabular}
\caption{Result per Feature}
\label{tab:features}
\end{table}

The best result is achieved using the features: \emph{user location, timezone and geoparsed user location}. Enabling all the features does not result in a higher accuracy. This is probably due to conflicting information of the features, such as the combination \emph{user language} and \emph{tweet language}.

\subsection{Error analysis}\label{sec:errorAnalysis}
By looking at the classification errors of the classifier we identified six common errors:

\begin{description}
\item[limited information] when the fields user location and timezone are empty the classifier cannot use the features \emph{timezone, user location, user location geocode, utc\_offset} and can only rely on \emph{tweet language}. Due to the large amount of tweets in the United States, according to the feature \emph{tweet language} English tweets are always in the United States.

\item[wrong timezone] In some tweets a timezone identifier does not correspond to the country. For example, the timezone may indicate Hawaii while the tweet is actually from Canada. 

\item[language detection twitter] The language detection of Twitter is limited in case of  short tweets. For example, the classifier classified a Dutch tweet with only the text ''feenstra'' (a Dutch family name) as a Swedish tweet.

\item[big classes] There are some features that can push the classifier into the wrong direction. An example is the class English for \emph{tweet language}. United States is the biggest class for English. When a tweet is in English and the location states India the tweet language class overrules the India class. Leading into a misclassification. The same applies to big classes of other features, such as timezone Pacific. 

\item[learning mistakes] The learner is limited to the provided training data. For example, a user from the Netherlands posted a dutch tweet in France. This causes the learner to associate meta-information of the Dutch user to France. However, when there are a many tweets to compensate the error of a Dutch tweet in France, the error will be neglect-able. 

\end{description}

In order to find out whether the results are consistent for each country, we execute a test for each country. We used all available geo information tweets from dataset \#1 as training data and we assumed that the distribution of the tweet meta data with geo-information, is the same as that of tweets without geo-information. For each country we checked whether our model matched the geo-information. For example, we have a list Dutch tweets (geo location in Netherlands) and we checked in how many cases our Bayes classifier estimated that the tweets are from the Netherlands, without looking at the geo location data. In table~\ref{tab:errors} (page \pageref{tab:errors}) the results are listed for each country. We omitted countries with less than 15 geo location tweets. The feature \emph{tweet language} caused in 36 of the 69 countries a decrease in accuracy. This can be due to the noise that \emph{tweet language} introduces, because some countries are sharing the same language. The effect on countries known to share the language, such as United Kingdom and United States and The Netherlands and Belgium, is negligible. For the Netherlands the accuracy increased with 8\% by adding \emph{tweet language}. For Belgium the accuracy decreased with 0.5\%.

To observe the effect of noise, we limited the classes countries to Europe. All other countries were classified as ''Other''. The accuracy went up to 88.15\%, due to the elimination of noise. Limiting the amount of countries did improve the accuracy.

\section{Discussion}
As shown above using dataset \#1 resulted in an accuracy of 82\%. However, dataset \#1 are random tweets with \emph{movember}. Therefore, there are some concerns about the data. The dataset is not uniform. There are a lot of tweets in English. This causes classification problem \textbf{big classes} (see page~\pageref{sec:errorAnalysis}) with the classifier. The accuracy might be lower using a uniform distributed dataset.

The dataset we have used is also small, only 40k tweets with geolocation. This causes the problem \textbf{learning mistakes}. This error can be reduced by having more training data. Some values are not known by the Bayes model and with more training data it might improve the accuracy.

The Bayes classifier is easy to use, however when using the \emph{user location} feature there are a lot of unknown locations. The classifiers can only handle known locations, for example New York or Amsterdam. Other classification engines can improve the accuracy. For example, decision tree~\cite{aggarwal2012survey} can be useful in this research due to the limited options of timezone and user language. 

\begin{table}[htbp]
\centering
\footnotesize
\begin{tabular}[c]{|l|l|p{1.5cm}|p{1.4cm}|p{1.5cm}|}
\hline \textbf{Country} & \textbf{\# tweets} & \textbf{Loc,~ TZ, geoparsed} & \textbf{Loc,~ TZ, language} & \textbf{Loc,~ TZ, language, geoparsed}  \\ \hline
United Kingdom & 15082 & 98.93 & 99.55 & 99.24\\ \hline 
United States & 9605 & 93.14 & 93.13 & 93.10\\ \hline 
Canada & 4603 & 79.62 & 82.40 & 78.51\\ \hline 
Netherlands & 877 & 79.02 & 72.98 & 87.12\\ \hline  
Ireland & 869 & 83.89 & 62.03 & 78.37\\ \hline 
Australia & 739 & 86.74 & 71.58 & 81.33\\ \hline 
South Africa & 649 & 90.29 & 82.59 & 87.83\\ \hline 
Spain & 648 & 85.80 & 85.96 & 89.20\\ \hline 
France & 619 & 79.81 & 83.04 & 85.95\\ \hline 
Indonesia & 581 & 87.95 & 79.17 & 84.68\\ \hline 
Norway & 397 & 80.60 & 84.38 & 88.16\\ \hline 
Coral Sea Islands & 351 & 48.15 & 44.16 & 46.72\\ \hline 
Sweden & 346 & 81.50 & 86.71 & 89.31\\ \hline 
Germany & 303 & 82.84 & 82.18 & 80.53\\ \hline 
Belgium & 293 & 87.71 & 86.35 & 87.37\\ \hline 
Italy & 259 & 73.75 & 74.90 & 78.38\\ \hline 
Finland & 245 & 84.49 & 93.06 & 91.02\\ \hline 
Denmark & 165 & 83.64 & 80.61 & 80.61\\ \hline 
Mexico & 150 & 75.33 & 58.00 & 70.00\\ \hline 
New Zealand & 143 & 83.92 & 69.93 & 81.12\\ \hline 
Malaysia & 142 & 85.92 & 79.58 & 80.28\\ \hline 
Brazil & 130 & 81.54 & 85.38 & 85.38\\ \hline 
Switzerland & 124 & 83.87 & 50.00 & 73.39\\ \hline 
Costa Rica & 119 & 66.39 & 63.87 & 66.39\\ \hline 
United Arab Emirates & 115 & 82.61 & 73.91 & 74.78\\ \hline 
India & 107 & 85.05 & 72.90 & 79.44\\ \hline 
Russia & 91 & 81.32 & 83.52 & 82.42\\ \hline 
Singapore & 81 & 77.78 & 60.49 & 60.49\\ \hline 
Philippines & 81 & 83.95 & 70.37 & 80.25\\ \hline 
Saudi Arabia & 73 & 86.30 & 72.60 & 76.71\\ \hline 
Hong Kong & 67 & 70.15 & 79.10 & 65.67\\ \hline 
Kuwait & 62 & 85.48 & 82.26 & 79.03\\ \hline 
Czech Republic & 62 & 79.03 & 69.35 & 72.58\\ \hline 
Austria & 61 & 81.97 & 77.05 & 81.97\\ \hline 
Egypt & 56 & 94.64 & 87.50 & 91.07\\ \hline 
Portugal & 49 & 85.71 & 73.47 & 81.63\\ \hline 
Japan & 46 & 82.61 & 80.43 & 82.61\\ \hline 
Latvia & 44 & 97.73 & 100.00 & 100.00\\ \hline 
Poland & 44 & 79.55 & 72.73 & 81.82\\ \hline 
Qatar & 34 & 79.41 & 76.47 & 79.41\\ \hline 
Chile & 32 & 87.50 & 81.25 & 84.38\\ \hline 
Turkey & 30 & 66.67 & 60.00 & 66.67\\ \hline 
Argentina & 27 & 70.37 & 74.07 & 70.37\\ \hline 
Greece & 24 & 70.83 & 70.83 & 79.17\\ \hline 
Puerto Rico & 24 & 79.17 & 70.83 & 75.00\\ \hline 
Kenya & 23 & 95.65 & 78.26 & 91.30\\ \hline 
Colombia & 23 & 60.87 & 30.43 & 30.43\\ \hline 
Hungary & 23 & 78.26 & 60.87 & 73.91\\ \hline 
Thailand & 22 & 36.36 & 36.36 & 36.36\\ \hline 
Jersey & 20 & 60.00 & 45.00 & 50.00\\ \hline 
Venezuela & 19 & 100.00 & 100.00 & 100.00\\ \hline 
Gibraltar & 17 & 94.12 & 94.12 & 94.12\\ \hline 
Isle of Man & 16 & 75.00 & 0.00 & 75.00\\ \hline 
Dominican Republic & 16 & 93.75 & 93.75 & 93.75\\ \hline 
South Korea & 15 & 46.67 & 46.67 & 40.00\\ \hline 
 & & & & \\ \hline
Average & & 77.84\% & 70.07\% & 74.94\% \\ \hline
Standard deviation & & 12.24 & 15.96 & 13.94 \\ \hline
Europe & & 88.15\% & 82.29\% & 87.83\% \\ \hline
\end{tabular}
\caption{Analysis of accuracy per country.}
\label{tab:errors}
\end{table}

\section{Conclusion}
Our initial goal was which features are required to identify the country of a single Tweet. Using meta-information of a tweet the location can identified. The Bayes classifier did perform well, with an accuracy of 82\% on the biggest dataset. Several features were tested and the combination of \emph{timezone, user location and geoparsed location} resulted in the highest accuracy. The feature \emph{user language} and \emph{tweet language} did improve for some countries the accuracy. The feature \emph{utc\_offset} did not improve the classifier and should not be used. 

The amount of meta-information is sufficient to determine the country of a single tweet. The features \emph{timezone} and \emph{geoparsed location} are the most important features and the combination results already in a classification accuracy of 73\%. However, the feature \emph{location} and \emph{tweet language} provided extra information, still improving the accuracy. 

Given the limitations our method performs well when a country has unique meta data, such as \emph{timezone} Amsterdam and \emph{user language} Dutch. However, our method underperforms for countries that share many properties such as Canada and United States (see Table~\ref{tab:errors}).

This method might be applicable on other social networks as well. Facebook and LinkedIn have similar information available through an API. The location of a message can be determined using a machine learning method, such as Na\"ive Bayes. This method can be applied to filter on countries and to clean up a Twitter stream. 

\section{Future work}
To improve this method the following features can be added or improved. In 25\% of the tweets the features \emph{timezone} and \emph{location} are empty. In those cases a feature using the tweet text might be useful. The tweeted text can contain places, which can be extracted using Entity Recognition. Also the information in the \emph{user description} can used for extraction locations.

Currently we are examine
 one tweet at a time, this is fast. However, there is more information available by looking at all the tweets of a user. When at tweet is retweeted the original tweet is embedded into the tweet. The user of the retweeted tweet might relate to the user location of who the tweet retweeted. 

Also a hashtag can contain useful information. For example, hashtag of a certain event can be used to detect where the Tweet is from. Using information related to the hashtag can improve the classifier.

\bibliographystyle{abbrv}

\end{document}